\documentclass{npw}
\usepackage{graphicx}
\usepackage{amsmath}
\usepackage{url}
\usepackage{pstricks}
\newcommand{\bk}{\boldsymbol{\mathbf k}}
\newcommand{\bkp}{\boldsymbol{\mathbf k^\prime}}
\newcommand{\vsigma}{\boldsymbol{\mathbf\sigma}}
\newcommand{\br}{\boldsymbol{\mathbf r}} 
\newcommand{\dd}{\delta(\br)}

\begin{document}

\title{Partial wave decomposition of the N3LO equation of state}
\author{
  D. Davesne\email{davesne@ipnl.in2p3.fr}, Jacques Meyer \email{jmeyer@ipnl.in2p3.fr} \\
  \it Universit\'e de Lyon, F-69003 Lyon, France; \\
  \it Universit\'e Lyon 1,
      43 Bd. du 11 Novembre 1918, F-69622 Villeurbanne cedex, France\\
  \it CNRS-IN2P3, UMR 5822, Institut de Physique Nucl{\'e}aire de Lyon \\
  A.Pastore\email{apastore@ulb.ac.be}\\
   \it Universit\'e Libre de Bruxelles, Physique Nucl\'eaire Th\'eorique, \\
   \it CP229, BE-1050 Bruxelles, Belgium
\\
   J. Navarro\email{navarro@ific.uv.es}\\
  \it IFIC (CSIC-Universidad de Valencia), Apartado Postal 22085, E-46.071-Valencia,
Spain   }
\date{\today}
\maketitle


\begin{abstract}
By means of a partial wave decomposition, we  separate their contributions to the equation of state of symmetric nuclear matter for the N3LO pseudo-potential. In particular, we show that although both the tensor and the spin-orbit terms do not contribute to the equation of state, they give a non-vanishing contribution to the separate $(JLS)$ channels. 
\end{abstract}


\section{Introduction}

The development of an universal nuclear energy density functional (NEDF)  represents an important goal in low-energy nuclear physics research. The NEDF represents the tool of choice for the investigation of static and dynamic properties in the region of medium to heavy mass nuclei from drip-line to drip-line~\cite{ben03}. A very extensive and detailed investigation on the properties of Skyrme functionals has been one the major objectives of the UNEDF-SciDAC collaboration~\cite{fur11,bog13}. It has been shown~\cite{kor13} that the \emph{standard} form of the Skyrme functional~\cite{per04} is not flexible enough to allow for further improvements of its spectroscopic qualities. Two possibilities are available, following either (i) the density functional theory, where the primary building block is the functional that includes all correlation effects, or (ii) the self-consistent mean-field theory, where the major ingredient is an effective pseudo-potential and correlations are added afterwards using a multi-reference approach~\cite{rin80}. Within the second approach, it is possible to add  correlations following a precise hierarchy towards the exact many-body ground state~\cite{yan07}.

Several groups have investigated the possibility of extending the \emph{standard} Skyrme functional by adding extra terms. Lesinski \emph{et al.}~\cite{les07} have studied the inclusion of a tensor term; Sadoudi \emph{et al.}~\cite{sad13} have derived a zero-range three-body term, and Carlsson \emph{et al.}~\cite{car08} have analyzed the contribution of higher order derivative terms to the functional.
However, the presence of finite-size instabilities related to the gradient terms of the Skyrme functional~\cite{les06} has made unpractical the task of fitting higher order terms. Recently, Hellemans \emph{et al.}~\cite{hel13} have presented a simple criterion based on Random Phase Approximation (RPA) calculations of homogenous symmetric nuclear matter (SNM) ~\cite{pas12} to detect and avoid these unphysical pathologies. In Ref.~\cite{pas13c}, Pastore \emph{et al.} have shown that combining the formalism of the linear response (LR) theory in SNM within the optimization procedure of a Skyrme functional, it is possible to obtain a functional free from this kind of pathologies. In Ref.~\cite{bec14}, we have presented the LR formalism for the N2LO Skyrme pseudo-potential~\cite{rai11,dav13} and we have also given the relevant expressions in spherical coordinates to solve the Hartree-Fock (HF) equations in spherical symmetry, thus presenting all the required ingredients to perform a fit of the parameters. The FIDIPRO group~\cite{fidipro} has published the numerical code HOSPHE~\cite{car10B}, which solves Hartree-Fock (HF) equations on an Harmonic Oscillator basis in spherical symmetry for the most general N3LO Skyrme functional~\cite{car08}.

Although all the necessary theoretical and numerical tools are available to perform a fit of the generalized Skyrme pseudo-potential, a crucial element is still missing, namely how to determine the set of observables or pseudo-observables that could optimally constrain the new terms. The study of the optimal merit function to be used into a fitting procedure is now the subject of an important debate within the nuclear structure community~\cite{dob14}. In fact, it would be preferable to identify for each term of the functional to identify the best set of observables which could better constrain it.

In the present article, we focus on the role of the constraints arising from infinite nuclear matter on the parameters of the functional. Usually this means adding a constraint on the values of the binding energy per particle, of saturation density and compressibility of SNM and also for pure neutron matter (PNM), as done for example for the Saclay-Lyon interactions~\cite{cha97}.
These informations are usually extracted from microscopic calculations based on realistic two- and three-body interactions as Brueckner-Hartree-Fock (BHF)~\cite{sch95}, self-consistent Green's functions (SCGF)~\cite{rio09,dic04}, auxiliary field diffusion Monte Carlo (AFDMC)~\cite{gan07}, Fermi hypernetted chain (FHNC)~\cite{pan79} or chiral effective field theory (CEFT), with renormalization group (RG)-evolved interactions constrained by nucleon scattering data~\cite{bog10}. It is worth mentioning that some of these calculations gives a more complete set of informations as for example the contribution to the total binding energy of each partial wave. This additional information is usually neglected since the \emph{standard} Skyrme functional contains only S and P wave terms and thus their structure is not rich enough to properly take into account these data, with the result that the parameters could be over-constrained. In Ref.~\cite{les06}, Lesinski \emph{et al.} have underlined that it is almost impossible to reproduce the contribution of the four spin/isospin channels in SNM to the total equation of state (EoS) based on a \emph{standard} Skyrme functional. The authors have also found that including additional density dependent terms does not improve the results. Similar conclusion can be drawn in Ref.~\cite{gor13}, where a different density dependency have been taken into account.

In this article, we determine and analyze the partial wave decomposition of the N3LO Skyrme pseudo-potential~\cite{rai11,dav14c},  and show that the inclusion of extra derivative term could be constrained by using \emph{ab-initio} results.
The article is organized as follows: in Sec.~\ref{sec:n2lo} we briefly present the N3LO Skyrme pseudo-potential in cartesian basis, while in 
Sec.~\ref{sec:partwaves} we illustrate the necessary formalism to derive the EoS in the different partial waves. Finally in Sec.\ref{sec:concl} we present our conclusions and perspectives.


\section{N3LO pseudo-potential}\label{sec:n2lo}

We write the N3LO Skyrme pseudo-potential as
\begin{eqnarray}\label{pseudo:skyrme}
v_{Sk}=v_{C}+v_{T}+v_{LS}+v_{3b}.
\end{eqnarray}
The central term $v_{C}$ can be written order by order as~\cite{dav14c}
\begin{eqnarray}\label{eq:vcen}
v_{C}=v^{(0)}+v^{(2)}+v^{(4)}+v^{(6)}
\end{eqnarray}
with
\begin{eqnarray}
v^{(0)}(\br) &=&  t_0 \, ( 1 + x_0 P_{\sigma} ) \,, \\
 v^{(2)}(\br) &=&  \tfrac{1}{2} \, t_1 \, ( 1 + x_1  P_{\sigma} )
      \big[ \bk^{\prime 2}  + \bk^2 \big]         +  t_2 \, ( 1 + x_2P_{\sigma}) \, \bkp \cdot  \bk    \, ,\\
v ^{(4)}(\br)  &=&   \frac{1}{4} t_1^{(4)} (1+x_1^{(4)} P_{\sigma}) \left[({\bf k}^2 + {\bf k'}^2)^2 + 4 ({\bf k'} \cdot {\bf k})^2\right] \nonumber\\
 & & + t_2^{(4)} (1+x_2^{(4)} P_{\sigma}) ({\bf k'} \cdot {\bf k}) ({\bf k}^2 + {\bf k'}^2) \label{V-four}\,,\\
 v ^{(6)}(\br)  &=& \frac{t_{1}^{(6)}}{2}(1+x_1^{(6)} P_{\sigma}) ({\bf k}^{'2} + {\bf k}^{2})\left[ ({\bf k}^{'2}+{\bf k}^{2})^{2}+12({\bf k}'\cdot{\bf k})^{2}\right] \nonumber\\
 & & + t_2^{(6)} (1+x_2^{(6)} P_{\sigma}) ({\bf k'} \cdot {\bf k}) \left[ 3({\bf k}^{'2}+{\bf k}^{2})^{2}+4({\bf k}'\cdot{\bf k})^{2}\right]\,, \nonumber\\
\end{eqnarray}
where $P_{\sigma}$ is the spin-exchange operator and a $\delta(\br)$ is implicit.
{The definitions of $\mathbf{r}$, $\mathbf{R}$, $\bk$, $\bkp$ are standard and can be found in the review paper of Bender \emph{et al.}~\cite{ben03}. }

\noindent Due to gauge-invariance, the spin-orbit term $v^{(LS)}$ is not modified by the inclusion of higher order terms~\cite{dav13} and it has the usual form
\begin{eqnarray}
\label{eq:Skyrme:LS}
v_\mathrm{LS} (\br) & = & i \, W_0 \, ( \vsigma_1 + \vsigma_2 ) \cdot \left[ \bk^{\prime} \times \dd \; \bk \right]\,.
\end{eqnarray}
The tensor term reads
\begin{eqnarray}
v_{T}&=&\frac{1}{2}t_{e}T_{e}(\mathbf{k}',\mathbf{k})+\frac{1}{2}t_{o}T_{o}(\mathbf{k}',\mathbf{k})\nonumber\\
&+&t_{e}^{(4)}\left[ (\mathbf{k}^{2}+\mathbf{k}^{'2})T_{e}(\mathbf{k}',\mathbf{k})+2(\mathbf{k}'\cdot\mathbf{k})T_{o}(\mathbf{k}',\mathbf{k})\right]\nonumber\\
&+&t_{o}^{(4)}\left[ (\mathbf{k}^{2}+\mathbf{k}^{'2})T_{o}(\mathbf{k}',\mathbf{k})+2(\mathbf{k}'\cdot\mathbf{k})T_{e}(\mathbf{k}',\mathbf{k})\right]\nonumber\\
&+&t_{e}^{(6)}\left[ \left(\frac{1}{4}(\mathbf{k}^{2}+\mathbf{k}^{'2})^{2}+(\mathbf{k}'\cdot\mathbf{k})^{2}\right)T_{e}(\mathbf{k}',\mathbf{k})+(\mathbf{k}^{2}+\mathbf{k}^{'2})(\mathbf{k}'\cdot\mathbf{k})T_{o}(\mathbf{k}',\mathbf{k})\right]\nonumber\\
&+&t_{o}^{(6)}\left[ \left(\frac{1}{4}(\mathbf{k}^{2}+\mathbf{k}^{'2})^{2}+(\mathbf{k}'\cdot\mathbf{k})^{2}\right)T_{o}(\mathbf{k}',\mathbf{k})+(\mathbf{k}^{2}+\mathbf{k}^{'2})(\mathbf{k}'\cdot\mathbf{k})T_{e}(\mathbf{k}',\mathbf{k})\right]\,,
\end{eqnarray}
where the operators $T_{e}$ and $T_{o}$ are defined as~\cite{dav14c}
\begin{eqnarray}\label{eq:teto}
T_{e}(\mathbf{k}',\mathbf{k})&=&3(\vsigma_{1}\cdot\mathbf{k}')(\vsigma_{2}\cdot\mathbf{k}')+3(\vsigma_{1}\cdot\mathbf{k})(\vsigma_{2}\cdot\mathbf{k})-(\mathbf{k}^{'2}+\mathbf{k}^{2})(\vsigma_{1}\vsigma_{2}),\\
T_{o}(\mathbf{k}',\mathbf{k})&=&3(\vsigma_{1}\cdot\mathbf{k}')(\vsigma_{2}\cdot\mathbf{k})+3(\vsigma_{1}\cdot\mathbf{k})(\vsigma_{2}\cdot\mathbf{k}')-2(\mathbf{k}'\mathbf{k})(\vsigma_{1}\vsigma_{2}).
\end{eqnarray}

\noindent Finally the three-body term is replaced in this paper by the usual density-dependent term~\cite{vau72}
\begin{eqnarray}\label{dd:term}
v_{3b}&\approx&\frac{1}{6}t_{3}(1+x_{3}P_{\sigma})\rho(\mathbf{R})^{\alpha}\delta(\mathbf{r})\,.
\end{eqnarray}

\noindent This extension to higher order derivative terms can be considered complementary to the analysis done in Ref.~\cite{sad13}. In fact, the pseudo-potential given in Eq.(\ref{pseudo:skyrme}) with the density dependent term given in Eq.~(\ref{dd:term}) will suffer from the same pathologies in multi-reference calculations as \emph{standard} Skyrme functionals~\cite{ben09}. We thus expect  that the next generation of Skyrme functionals should include higher order derivatives and explicit three-body terms (at least for the central part of the potential). This interaction would be free from the drawbacks detected in Ref.~\cite{ben09} since it would be a real Hamiltonian, which implies that the restoration of broken symmetries trough projection techniques~\cite{ben03} will be less problematic. Since we limit ourselves to single-reference calculations, we keep in this paper the density-dependent term to simplify our calculations.

We finally remind that the parameters used here to define the N3LO in Cartesian basis can be easily expressed in terms of the notation adopted in Ref.~\cite{rai11}. Explicit expressions can be found in Refs.~\cite{dav13,dav14c}.

\section{Partial wave decomposition}\label{sec:partwaves}

The partial wave decomposition is a very useful tool since it allows us to properly identify the contributions of the different terms to the total equation of state.
This analysis has been inspired by the previous results obtained by Baldo~\emph{et al.}~\cite{bal97}. By means of BHF methods, they have calculated the potential energy in different $(JLS)$-channels, where $\vec{J}=\vec{L}+\vec{S}$ is the total angular momentum, while $L,S$ represents the total orbital angular momentum and spin. 
Although such approach is not strictly related to the N3LO pseudo-potential, it has never been employed systematically to analyze Skyrme functionals.

Within the HF approximation, the potential energy for the central terms can be written with the usual spectroscopic notation as
\begin{equation}\label{eq:gsenergy1}
E_p(^{2S+1}L_J) = (2J+1) (2S+1)(2T+1) V(^{(2S+1)}L) \,, \\
\end{equation}
where $T$ is the total isospin quantum number and
\begin{eqnarray}\label{eq:gsenergy2}
V(^{2S+1}L) &=& \frac{1}{2}\sum_{ij}\langle ij|v^{(L)}(1-P_{x}P_{\sigma}P_{\tau})|ij\rangle .
\end{eqnarray}
The exchange term is explicitly included through the operator product $P_{x}P_{\sigma}P_{\tau}$, of the space exchange (Majorana), spin and isospin operators. For a contact interaction, $P_{x}$ can be replaced by $(-1)^{L}$so that a selection rule $L+S+T=\text{odd}$ arises from the product $P_{x}P_{\sigma}P_{\tau}$. The value of $T$ is therefore fixed by the values of $L$ and $S$.

As is well-known, the \emph{standard} Skyrme functional contains contributions only from S- and P- waves. The higher order terms introduce also a D- and F- wave, thus leaving more flexibility for comparisons with \emph{ab-initio} methods. To identify the contributions of a given pseudo-potential term to the different partial waves we simply expand it in spherical harmonics as 
 
\begin{eqnarray}
F(\mathbf{k}',\mathbf{k})&=&\sum_{L'M_{L}'LM_{L}}F_{L'M_{L}';LM_{L}}(k',k)Y^{*}_{L'M_{L}'}(\hat{k}')Y_{LM_{L}}(\hat{k})\,,
\label{eq:decomp}
\end{eqnarray}
with
\begin{eqnarray}
F_{L'M_{L}';LM_{L}}(k',k)&=&\int d\hat{k}' d \hat{k}Y_{L'M_{L}'}(\hat{k}')Y^{*}_{LM_{L}}(\hat{k})F(\mathbf{k}',\mathbf{k})\,.
\end{eqnarray} 
To help in the identification, we have associated indices $L'$ and $M_L'$ to momentum  ${\mathbf k'}$.

 \subsection{Central term}
For both the central and density-dependent terms, it is simple to show that the only non-vanishing terms are those with $L=L'$ and $M_{L}=M_{L}'$. The decomposition (\ref{eq:decomp}) can then be easily performed. However, the contribution for the different channels can also be identified directly by using the relation between Legendre polynomials of argument $(\hat{\mathbf k'} \cdot \hat{\mathbf k})$ and spherical harmonics of arguments  $\hat{\mathbf k'}$ and $ \hat{\mathbf k}$. We can then rewrite Eq.~(\ref{eq:vcen}) as

\begin{equation}
v_{C}=v^{(S)}+v^{(P)}+v^{(D)}+v^{(F)}\,,
\end{equation}
with
\begin{eqnarray}\label{eq:swave}
v^{(S)}&=&t_{0}(1+x_{0}P_{\sigma})+\frac{1}{6}t_{3}(1+x_{3}P_{\sigma})\rho^{\alpha}+\frac{1}{2}t_{1}(1+x_{1}P_{\sigma})({\bf k}^{'2}+{\bf k}^{2})\nonumber\\
&+&\frac{1}{4}t_{1}^{(4)}(1+x_{1}^{(4)}P_{\sigma})\left[({\bf k}^{'2}+{\bf k}^{2})^{2}+\frac{4}{3}\mathbf{k}^{'2}\mathbf{k}^{2} \right]\nonumber\\
&+&\frac{1}{2}t_{1}^{(6)}(1+x_{1}^{(6)}P_{\sigma})(\mathbf{k}^{2}+\mathbf{k}^{'2})\left[({\bf k}^{'2}+{\bf k}^{2})^{2}+4\mathbf{k}^{'2}\mathbf{k}^{2} \right]\,,\\
\label{eq:pwave}
v^{(P)}&=&t_{2}(1+x_{2}P_{\sigma})({\bf k}^{'}\cdot{\bf k}^{})\nonumber\\
&+&t_{2}^{(4)}(1+x_{2}^{(4)}P_{\sigma})({\bf k}^{'}\cdot{\bf k}^{}) ({\bf k}^{'2}+{\bf k}^{2})^{2}\nonumber\\
&+&t_{2}^{(6)}(1+x_{2}^{(6)}P_{\sigma})({\bf k}^{'}\cdot{\bf k}^{}) \left[3({\bf k}^{'2}+{\bf k}^{2})^{2}+\frac{12}{5}\mathbf{k}^{'2}\mathbf{k}^{2} \right]\,,\\
\label{eq:dwave}
v^{(D)}&=&\frac{1}{4}t_{1}^{(4)}(1+x_{1}^{(4)}P_{\sigma})\frac{4}{3} \left[3({\bf k}^{'}\cdot{\bf k}^{})^{2}-\mathbf{k}^{'2}\mathbf{k}^{2} \right] \nonumber\\
&+&\frac{1}{2}t_{1}^{(6)}(1+x_{1}^{(6)}P_{\sigma})4 ({\bf k}^{'2}+{\bf k}^{2})\left[ 3({\bf k}^{'}\cdot{\bf k}^{})^{2}-\mathbf{k}^{'2}\mathbf{k}^{2} \right]\,,
\end{eqnarray}
and
\begin{eqnarray}\label{eq:fwave}
v^{(F)}&=&t_{2}^{(6)}(1+x_{2}^{(6)}P_{\sigma})\frac{4}{5} ({\bf k}^{'}\cdot{\bf k}^{})\left[ 5({\bf k}^{'}\cdot{\bf k}^{})^{2}-3\mathbf{k}^{'2}\mathbf{k}^{2} \right]\,.
\end{eqnarray}

By inserting the Eqs.~(\ref{eq:swave}-\ref{eq:fwave}) in Eqs.~(\ref{eq:gsenergy1}-\ref{eq:gsenergy2}), we obtain the following explicit expressions 
\begin{eqnarray}
\frac{1}{A}E_p(^{1}S_0)&=&\frac{3}{16}t_{0}(1-x_{0})\rho+\frac{1}{32}t_{3}(1-x_{3})\rho^{\alpha+1}+\frac{9}{160}t_{1}(1-x_{1})\rho k_{F}^{2}\nonumber\\
&+&\frac{9}{280}t_{1}^{(4)}(1-x_{1}^{(4)})\rho k_{F}^{4}+\frac{1}{10}t_{1}^{(6)}(1-x_{1}^{(6)})\rho k_{F}^{6}\,,\\
\frac{1}{A}E_p(^{3}S_1)&=&\frac{3}{16}t_{0}(1+x_{0})\rho+\frac{1}{32}t_{3}(1+x_{3})\rho^{\alpha+1}+\frac{9}{160}t_{1}(1+x_{1})\rho k_{F}^{2}\nonumber\\
&+&\frac{9}{280}t_{1}^{(4)}(1+x_{1}^{(4)})\rho k_{F}^{4}+\frac{1}{10}t_{1}^{(6)}(1+x_{1}^{(6)})\rho k_{F}^{6}\,,
\end{eqnarray}
\begin{eqnarray}
\frac{1}{A}E_p(^{1}P_0)&=&\frac{3}{160}t_{2}(1-x_{2})\rho k_{F}^{2}+\frac{9}{560}t_{2}^{(4)}(1-x_{2}^{(4)})\rho k_{F}^{4}+\frac{3}{50}t_{2}^{(6)}(1-x_{2}^{(6)})\rho k_{F}^{6}\,,\\
\frac{1}{A}E_p(^{3}P_0)&=&\frac{3}{160}t_{2}(1+x_{2})\rho k_{F}^{2}+\frac{9}{560}t_{2}^{(4)}(1+x_{2}^{(4)})\rho k_{F}^{4}+\frac{3}{50}t_{2}^{(6)}(1-x_{2}^{(6)})\rho k_{F}^{6}\,,\\
\frac{1}{A}E_p(^{3}P_1)&=&\frac{9}{160}t_{2}(1+x_{2})\rho k_{F}^{2}+\frac{27}{560}t_{2}^{(4)}(1+x_{2}^{(4)})\rho k_{F}^{4}+\frac{9}{50}t_{2}^{(6)}(1-x_{2}^{(6)})\rho k_{F}^{6}\,,\\
\frac{1}{A}E_p(^{3}P_2)&=&\frac{3}{32}t_{2}(1+x_{2})\rho k_{F}^{2}+\frac{9}{112}t_{2}^{(4)}(1+x_{2}^{(4)})\rho k_{F}^{4}+\frac{3}{10}t_{2}^{(6)}(1-x_{2}^{(6)})\rho k_{F}^{6}\,,
\end{eqnarray}
\begin{eqnarray}
\frac{1}{A}E_p(^{1}D_2)&=&\frac{9}{560}t_{1}^{(4)}(1-x_{1}^{(4)})\rho k_{F}^{4}+\frac{1}{10}t_{1}^{(6)}(1-x_{1}^{(6)})\rho k_{F}^{6}\,,\\
\frac{1}{A}E_p(^{3}D_1)&=&\frac{9}{2800}t_{1}^{(4)}(1+x_{1}^{(4)})\rho k_{F}^{4}+\frac{1}{50}t_{1}^{(6)}(1+x_{1}^{(6)})\rho k_{F}^{6}\,,\\
\frac{1}{A}E_p(^{3}D_2)&=&\frac{3}{560}t_{1}^{(4)}(1+x_{1}^{(4)})\rho k_{F}^{4}+\frac{1}{30}t_{1}^{(6)}(1+x_{1}^{(6)})\rho k_{F}^{6}\,,\\
\frac{1}{A}E_p(^{3}D_3)&=&\frac{3}{400}t_{1}^{(4)}(1+x_{1}^{(4)})\rho k_{F}^{4}+\frac{7}{150}t_{1}^{(6)}(1+x_{1}^{(6)})\rho k_{F}^{6}\,,
\end{eqnarray}
\begin{eqnarray}
\frac{1}{A}E_p(^{1}F_3)&=&\frac{1}{150}t_{2}^{(6)}(1-x_{2}^{(6)})\rho k_{F}^{6}\,,\\
\frac{1}{A}E_p(^{3}F_2)&=&\frac{1}{70}t_{2}^{(6)}(1+x_{2}^{(6)})\rho k_{F}^{6}\,,\\
\frac{1}{A}E_p(^{3}F_3)&=&\frac{1}{50}t_{2}^{(6)}(1+x_{2}^{(6)})\rho k_{F}^{6}\,,\\
\frac{1}{A}E_p(^{3}F_4)&=&\frac{9}{350}t_{2}^{(6)}(1+x_{2}^{(6)})\rho k_{F}^{6}\,.
\end{eqnarray}
 
\subsection{Tensor term}
 
Due to its spin structure, the tensor does not contribute to the global EoS. However, as we can see below, it should be realized that this vanishing contribution actually results from the sum of individual non-vanishing contributions in different channels $(JLST)$.
{In presence of the tensor term, $L$ is not a good quantum number, and we have to couple it to $S$ (which is always equal to 1, as it is explicitly written below) and write our expressions in terms of the total angular momentum $J$.} 

Both operators $T_{e}$ and $T_{o}$ are second-order rank tensors in the spin space. Consequently, they can only contribute to the triplet partial waves. In Appendix~{\ref{Te}, we show that all contributions proportional to $T_{e}$ are identically zero, so that we are left with the contributions coming from the $T_{o}$ operator only. Moreover, we show in  Appendix~\ref{To}, that the contributions from $T_{o}$ are non zero only between \emph{bra-kets} with the same orbital momentum $L=1$. After some tedious calculations, we find 
\begin{eqnarray}
\frac{1}{A}E_p(^{3}P_{0})&=&-\frac{3}{80}t_{o}\rho k_{F}^{2}-\frac{9}{140}t_{o}^{(4)}\rho k_{F}^{4}-\frac{17}{750}t_{0}^{(6)}\rho k_{F}^{6}\,,\\
\frac{1}{A}E_p(^{3}P_{1})&=&\frac{9}{160}t_{o}\rho k_{F}^{2}+\frac{27}{280}t_{o}^{(4)}\rho k_{F}^{4}+\frac{17}{500}t_{0}^{(6)}\rho k_{F}^{6}\,,\\
\frac{1}{A}E_p(^{3}P_{2})&=&-\frac{3}{160}t_{o}\rho k_{F}^{2}-\frac{9}{280}t_{o}^{(4)}\rho k_{F}^{4}-\frac{17}{1500}t_{0}^{(6)}\rho k_{F}^{6}\,,
\end{eqnarray}
\begin{eqnarray}
\frac{1}{A}E_p(^{3}D_{1})&=&-\frac{9}{1000}t_{e}^{(4)}\rho k_{F}^{4}-\frac{7}{1500}\rho k_{F}^{6}\,,\\
\frac{1}{A}E_p(^{3}D_{2})&=&\frac{3}{200}t_{e}^{(4)}\rho k_{F}^{4}+\frac{7}{900}\rho k_{F}^{6}\,,\\
\frac{1}{A}E_p(^{3}D_{3})&=&-\frac{3}{500}t_{e}^{(4)}\rho k_{F}^{4}-\frac{7}{2250}\rho k_{F}^{6}\,,
\end{eqnarray}
\begin{eqnarray}
\frac{1}{A}E_p(^{3}F_{2})&=&-\frac{3}{875}t_{0}^{(6)}\rho k_{F}^{6}\,,\\
\frac{1}{A}E_p(^{3}F_{3})&=&\frac{3}{500}t_{0}^{(6)}\rho k_{F}^{6}\,,\\
\frac{1}{A}E_p(^{3}F_{4})&=&-\frac{9}{3500}t_{0}^{(6)}\rho k_{F}^{6}\,.
\end{eqnarray}
The above formula constitute the first noticeable result of this paper. They show that we can have some constraints on the tensor parameters coming from the partial wave contributions to the equation of state by using $ab-initio$ results.
It is worth noticing that the tensor term lifts the degeneracy between the different partial waves, but for a given value of $L$ the sum of the different $J$ terms gives zero, as expected. 

\subsection{Spin-orbit term}
 
The spin-orbit term is not modified by the inclusion of higher order derivative terms. Contrary to the original intuition of Skyrme~\cite{bel56}, the term given in Eq.~(\ref{eq:Skyrme:LS}) is the only one which preserves gauge invariance~\cite{dob95}. Possible contributions to the spin-orbit could however arise from higher order tensor terms~\cite{rin80}. 
 
The particular structure of the Pauli matrices $(\vsigma_{1}+\vsigma_{2})$ implies that this term gives no contribution to the global EoS. However, like the tensor, it gives a non-vanishing contribution to the different $P$-waves. Explicitly, we have
\begin{eqnarray}
 \frac{1}{A}E_p(^{3}P_{0})&=&\frac{1}{40}W_{o}\rho k_{F}^{2}\,,\\
\frac{1}{A}E_p(^{3}P_{1})&=&\frac{3}{80}W_{o}\rho k_{F}^{2}\,,\\
\frac{1}{A}E_p(^{3}P_{2})&=&-\frac{1}{16}W_{o}\rho k_{F}^{2}\,.
\end{eqnarray}
One can directly check that the sum of these terms gives zero, as expected. Here again, we can see that the spin-orbit parameter is constrained by the partial wave contributions to the equations of state. 


\section{Conclusions and Perspectives}\label{sec:concl}

We have calculated for the first time the contributions to the EoS of the N3LO Skyrme pseudo-potential in the different $(JLST)$ channels. 
Although both the tensor and the spin-orbit terms do not contribute to the total EoS, we have show that they give a non-vanishing contribution to the separate $(JLST)$ channels. To our knowledge, there are no studies on the impact of the tensor and spin-orbit terms on the EoS for phenomenological functionals (not necessary zero-range one). It is thus mandatory to make a sensitivity analysis for the different partial waves obtained from different \emph{ab-initio} methods and different realistic two- and three-body nuclear interactions. The goal of such an analysis would be not to improve the quality of the global EoS for the Skyrme functional, but to use several pseudo-observables derived from microscopic calculations to introduce additional constraints to the parameters of the pseudo-potential. Furthermore, compared to previous attempts of determining the parameters of the tensor using finite nuclei observables~\cite{les07,cao11,bai11}, our equations have the advantage of not being polluted by finite size and shell effects. All these aspects are left for a forthcoming publication.

\begin{ack}
The work of J.N. has been supported by grant FIS2011-28617-C02-2, Mineco (Spain).
\end{ack}

\appendix
\section*{Appendix}

In the following we give some useful expressions used to manipulate the tensor $T_{e}$ and $T_{o}$ operators.
By using properties of Pauli matrices, it is possible to show that $Q_{T}P^{(S=0)}_{\sigma}=0$ and $Q_{T}P^{(S=1)}_{\sigma}=Q_{T}$; where $Q_{T}$ stands for a generic tensor operator ($i.e.$ $T_{e}$ or $T_{o}$).

\section{Contribution of the $T_{e}$ operator}\label{Te}

The tensor operator $T_{e}$ can be easily written in the spherical basis as
\begin{eqnarray}
T_{e}(\mathbf{k}',\mathbf{k})&=&8\pi \sqrt{\frac{2\pi}{15}}\sum_{\mu_{1}\mu_{2}} 
(1, 1, 2; \mu_1, \mu_2, \mu_1+\mu_2)
 [\sigma_{1}]_{1,\mu_{1}} [\sigma_{2}]_{1,\mu_{2}}\left[k'^{2}Y^*_{2,\mu_{1}+\mu_{2}}(\hat{k}')+k^{2}Y^*_{2,\mu_{1}+\mu_{2}}(\hat{k}) \right]\,,
\end{eqnarray}
where $(j_1, j_2, J; m_1, m_2, M)$ stands for a Clebsch-Gordan coefficient. 
This term can give only non-zero contributions between \emph{bra-kets} with $|L-L'|=2$. Moreover,
by performing the angular integral, one can immediately observe that the $T_{e}$ contribution vanishes.

\section{Contribution of the $T_{o}$}\label{To}

Similarly, the tensor operator $T_{o}$ can be written as
\begin{eqnarray}
T_{o}&=&4\pi k'k \sum_{\mu_{1}\mu_{2}} (-)^{\mu_{2}} Y^*_{1,\mu_{1}}(\hat{k}') Y_{1,\mu_{2}}(\hat{k})
\left\{ 
 + [\sigma_1]_{1,\mu_{1}} [\sigma_2]_{1,-\mu_{2}}+ [\sigma_1]_{1,-\mu_{2}} [\sigma_2]_{1,\mu_{1}} \right\} \nonumber \\
&& -\frac{8\pi}{3} (\vsigma_{1} \cdot \vsigma_{2}) k'k\sum_{\mu}Y_{1\mu_{}}^{*}(\hat{k}')Y_{1\mu_{}}(\hat{k})
\end{eqnarray}
from which we find that the only non vanishing matrix elements have $L=L'=1$. We moreover notice the useful following property
\begin{equation}
T_{o}(1-P_{x}P_{\sigma}P_{\tau})=2T_{o}\delta_{S1}\delta_{T1}\delta_{L1}.
\end{equation}
At fourth order, the  remaining tensor contributions have the form $f(\mathbf{k}',\mathbf{k})T_{o}$, where $f(\mathbf{k}',\mathbf{k})=\mathbf{k}'\cdot\mathbf{k} \text{ or } (\mathbf{k}^{'2}+\mathbf{k}^{2})$. The term $(\mathbf{k}^{'2}+\mathbf{k}^{2})$ does not change the angular structure so it will contribute to the same channel as the second order term, while for the $\mathbf{k}'\cdot\mathbf{k}$ term, one can show that the non-vanishing matrix elements have $L=2$ and thus $T=0$. The same applies to the sixth order : $f(\mathbf{k}',\mathbf{k})=(\mathbf{k}^{'2}+\mathbf{k}^{2})^{2}$ contributes in the $L=1,T=1$ channel, the term $f(\mathbf{k}',\mathbf{k})=(\mathbf{k}^{'2}+\mathbf{k}^{2})(\mathbf{k}'\cdot \mathbf{k})$ contributes to the $L=2,T=0$ channel and finally the $f(\mathbf{k}',\mathbf{k})=(\mathbf{k}'\cdot \mathbf{k})^{2}$ contributes to the $L=3,T=1$ and $L=1,T=1$ channels. 

\bibliography{biblio}

\end{document}